\documentclass[fleqn,usenatbib]{mnras}
\def\etal{{\it et\thinspace al.}\ }
\def\hb{{H\,$\beta$}\ }
\def\ha{{H\,$\alpha$}\ }

\def\hii{{{\sc H\,ii}}\ }
\def\mniii{{{\sc Mn\,iii}}\ }
\def\oii{{O~\sc ii}\ }
\def\sii{{S~\sc ii}\ }
\def\oiii{{O~\sc iii}\ }

\def\feii{{\rm Fe~\sc ii}\ }

\def\eion{{(e~+~ion)}\ }
\def\te{{T$_e$}}

\def\en{{$n$}\ }
\def\eg{{\it e.g.}\ }

\usepackage{newtxtext,newtxmath}
\usepackage{booktabs} 
\usepackage[T1]{fontenc}
\usepackage{subcaption}
\usepackage{placeins}
\DeclareRobustCommand{\VAN}[3]{#2}
\let\VANthebibliography\thebibliography
\def\thebibliography{\DeclareRobustCommand{\VAN}[3]{##3}\VANthebibliography}

\usepackage{graphicx}	% Including figure files
\usepackage{amsmath}	% Advanced maths commands
\usepackage{float}
\usepackage{booktabs}
\usepackage{siunitx}
\usepackage{longtable}
\usepackage{array}

\title[{[Mn III]} line ratios]{Emissivity line ratios for [Mn III] and spectral diagnostics of H II regions}

\author[Z. Samak et al.]{
Zher Samak,$^{1}$
Sultana N. Nahar,$^{2}$
Anil K. Pradhan$^{2,3}$
\\
$^{1}$Al Aqsa University, Gaza, Palestine\\
$^{2}$Department of Astronomy, The Ohio State University, Columbus, Ohio, USA 43210\\
$^{3}$Chemical Physics Program, The Ohio State University, Columbus, Ohio, USA 43210
}

\date{Accepted XXX. Received YYY; in original form ZZZ}

\pubyear{\the\year{}}

\begin{document}
\label{firstpage}
\pagerange{\pageref{firstpage}--\pageref{lastpage}}
\maketitle

\begin{abstract}
We present the first theoretical emission-line study of [\mniii] as potential determinant of physical conditions and manganese abundance.
We compute atomic data for a set of [\mniii] lines and line emissivity ratios as temperature-density diagnostics. Accurate relativistic Breit-Pauli R-matrix calculations have been
carried out for electron impact excitation of [\mniii] transitions
that have not been previously studied. Selected forbidden lines are shown to be sensitive
to temperature and density under nebular conditions in \hii regions such as SNRs.
The coupled channel R-matrix calculations include a wavefunction expansion that includes 38
levels of \mniii dominated by the ground and first excited electronic configurations $3d^5$ and
$3d^44s$.
Collision strengths for all 703 forbidden transitions among those levels, and Maxwellian averaged
rate coefficients are obtained. A collisional-radiative model yields  emissivity line ratios using
calculated collision strengths and radiative decay A-values. Based on the present analysis, certain [\mniii] lines are predicted to be sensitive
to density and/or in the typical nebular range. The detection of those lines could also yield cosmic Mn abundance relative to other elements.
\end{abstract}


%%%%%%%%%%%%%%%%%%%%%%%%%%%%%%%%%%%%%%%%%%%%%%%%%%

%%%%%%%%%%%%%%%%% BODY OF PAPER %%%%%%%%%%%%%%%%%%

Formation of Fe-group elements occurs via explosive supernovae nucleosynthesis (\eg \citealt{nissen2024,ting2022}). Of these elements, Fe and Mn can both be produced in core collapse supernovae (SNe) Type II with massive stars as progenitors at the end of stellar nucleosynthesis, and in SNe Type~Ia following a binary system explosion with white dwarf progenitors formed from low-mass stars. However, the timescales are quite different since massive stars have lifetimes of Myrs whereas white dwarf formation takes place at typical low-mass stellar lifetimes of Gyrs \citep{Jomaron_1999,Berg_2015}. Manganese is an odd-Z element and has a different temporal history than Fe.
The Mn/Fe ratio increases with time or with Fe/H as function of the dichotomy between massive core collapse Supernovae Type II (SNe II) and Supernovae Type Ia (SNe Ia). Produced more efficiently in the neutron-rich environment of SNe~Ia more than SNe~II, the Mn abundance relative to Fe continues to increase and is therefore a chronometer of galaxy evolution, nucleosynthesis, and star formation history of the Universe.
Abundance determinations are made from spectral lines. Generally, neutral Mn~I absorption lines are measured from stellar atmospheres relative to Fe I to infer the abundance ratio, such as in the Milky Way APOGEE survey \citep{Majewski2017,Smith_2021,Melo_2024}. 
 Whereas absorption by neutral Mn I is widely employed as a probe of Mn abundance through stellar spectral analysis, Mn ions have not been studied in emission. However, highly ionized Mn ions are detected in X-ray observations of 
supernova remnants. It is therefore expected that low ionization stages might be present in late nebular phases of SNRs, emitting in the UV/O/IR, and corresponding emission lines may provide useful information on physical conditions and abundances.

To our knowledge, emission lines from ionized Mn have not been employed for astrophysical diagnostics of physical conditions and elemental abundances from H~II regions such as supernova remnants (SNRs). The primary reasons are twofold: (i) Mn abundance is lower than Fe by about a factor of 100 and spectral lines are difficult to detect, and (ii) singly ionized Mn II lines are in the UV and not readily observed from ground-based surveys. However, red-shifted lines from high-$z$ objects are now well within the range of the JWST/NIRSpec and should be eminently observable.
In this work we focus on relatively low-lying energy levels of [Mn\,{\sc iii}] and NUV/O/NIR forbidden line transitions and emissivity ratios, that could also lead to determination of gas phase abundance of Mn relative to other elements. 
 
\vspace{-10pt}
\section{Theory and computations}
The atomic processes that give rise to emission lines considered in this work are briefly described below, in addition to the calculations and data obtained. 
\vspace{-10pt}
\subsection{Energy levels and radiative transitions}
The energy levels and Einstein $A$-coefficients for the forbidden $E2$ and 
$M1$ transitions among the fine-structure levels of the ground-state manifold in [Mn\,\textsc{iii}] were calculated using the \textsc{SUPERSTRUCTURE} (SS) code 
\citep{Eissner1974}. 
The configuration interaction (CI) calculations included 
spectroscopic levels of the ground configuration $3d^5$ and a few levels of $3d^4s$, together with 
\en = 4,5 correlation configurations: $3d^{4}\,(4s,4p,4d,4f,5s,5p,5d)$. 
A total of 1421 fine-structure levels were computed by SS, spanning an energy range from 
the ground state $^6S_{5/2}$ to approximately 2.2~Ry. Of 
these, the lowest 38 levels dominated by the $3d^5$ and $3d^44s$ were selected as target states for subsequent 
\eion scattering R-matrix calculations (described below).
The selected fine-structure levels of the are of even parity, and all transitions
among these occur via electric-quadrupole ($E2$) or magnetic-dipole ($M1$) relativistic operators.
The calculated energy levels were compared with the available 
experimental values in the NIST Atomic Spectra Database 
\citep{Kramida2023} to assess the reliability of atomic structure 
calculations prior to proceeding with calculations of transition probabilities and collision strengths. A comparison of the calculated SUPERSTRUCTURE energies with experimental values in the NIST Atomic Spectra Database 
\citep{Kramida2023} for the lowest 20 fine-structure levels 
(Table~1) shows a mean difference of about 3 per cent with respect 
to NIST, providing a benchmark for the expected accuracy of the 
present atomic structure calculations. General accuracy of 
R-matrix data is estimated to be approximately 10--15 per cent 
based on experimental benchmarking (e.g.\ \citealt{Sigut1995}, for 
Mg~II).\\
(see Table~\ref{tab:levels}).
\begin{table}
\caption{Fine-structure energy levels of [Mn\,{\sc iii}] from the 
ground configuration $3d^5$. Level indices follow the NIST ordering. 
Energies are given in Rydberg (Ry). SS denotes the present 
{\sc SUPERSTRUCTURE} calculations; NIST values are from 
\citet{Kramida2023}.}
\label{tab:levels}
\resizebox{\columnwidth}{!}{%
\begin{tabular}{clccrr}
\hline
Index & Configuration & Term & $J$ & $E_{\rm SS}$ (Ry) & $E_{\rm NIST}$ (Ry) \\
\hline
 1 & $3d^5$ & $^6S$   & $5/2$  & 0.0000000  & 0.0000000  \\
 2 & $3d^5$ & $^4G$   & $11/2$ & 0.2482025  & 0.2444419  \\
 3 & $3d^5$ & $^4G$   & $9/2$  & 0.2481890  & 0.2446852  \\
 4 & $3d^5$ & $^4G$   & $5/2$  & 0.2480204  & 0.2447463  \\
 5 & $3d^5$ & $^4G$   & $7/2$  & 0.2481159  & 0.2447654  \\
 6 & $3d^5$ & $^4P$   & $5/2$  & 0.2829707  & 0.2657956  \\
 7 & $3d^5$ & $^4P$   & $3/2$  & 0.2831493  & 0.2661565  \\
 8 & $3d^5$ & $^4P$   & $1/2$  & 0.2833889  & 0.2664672  \\
 9 & $3d^5$ & $^4D$   & $7/2$  & 0.3036693  & 0.2944058  \\
10 & $3d^5$ & $^4D$   & $1/2$  & 0.3037763  & 0.2949670  \\
11 & $3d^5$ & $^4D$   & $5/2$  & 0.3041189  & 0.2951020  \\
12 & $3d^5$ & $^4D$   & $3/2$  & 0.3040145  & 0.2951111  \\
13 & $3d^5$ & $^2I$   & $11/2$ & 0.3571703  & 0.3569830  \\
14 & $3d^5$ & $^2I$   & $13/2$ & 0.3573770  & 0.3570030  \\
15 & $3d^5$ & $^2D$   & $5/2$  & 0.3940965  & 0.3757890  \\
16 & $3d^5$ & $^2D$   & $3/2$  & 0.3959121  & 0.3788120  \\
17 & $3d^5$ & $^2F$   & $7/2$  & 0.4067494  & 0.3882590  \\
18 & $3d^5$ & $^2F$   & $5/2$  & 0.4060531  & 0.3928050  \\
19 & $3d^5$ & $^4F$   & $9/2$  & 0.4170512  & 0.3970678  \\
20 & $3d^5$ & $^4F$   & $7/2$  & 0.4171292  & 0.3973352  \\
\hline
\end{tabular}%
}
\end{table}
\suppressfloats[t] 
%\FloatBarrier
%\subsection{Radiative transitions}
 The forbidden $E2,M1$ line strengths and $A$-values are calculated using the SS code.
The $A$-values within the forbidden fine-structure transitions of the ground configuration complex
are generally small, in the range $10^{-3}$ to $10^{2}~\mathrm{s^{-1}}$. Table \ref{tab:transitions} gives a representative sample of the computed $A$-values that are employed in the line emissivity calculations.
%\begin{table}[H]
\begin{table}
\caption{Selected forbidden transitions among the lowest 20 
fine-structure levels of [Mn\,{\sc iii}] ($3d^5$ configuration).} 
\label{tab:transitions}
\centering
\small  % تصغير الخط قليلاً ليتسع الجدول
\begin{tabular}{r l l l l r r}
\hline
No. & \multicolumn{2}{c}{Upper level} & 
     \multicolumn{2}{c}{Lower level} & 
$A$ (s$^{-1}$) & $\lambda$ (\AA) \\
\cline{2-3}\cline{4-5}
& Term & $J$ & Term & $J$ & & \\
\hline
1 & $^{4}G$ & 5/2  & $^{6}S$ & 5/2  & 2.581($-$08) &  3674.16 \\
 2 & $^{4}G$ & 7/2  & $^{6}S$ & 5/2  & 8.384($-$10) &  3672.75 \\
 3 & $^{4}G$ & 9/2  & $^{6}S$ & 5/2  & 4.447($-$12) &  3671.67 \\
 4 & $^{4}P$ & 5/2  & $^{6}S$ & 5/2  & 2.989($-$01) &  3220.36 \\
 5 & $^{4}P$ & 5/2  & $^{4}G$ & 5/2  & 2.797($-$06) & 26073.21 \\
 6 & $^{4}P$ & 5/2  & $^{4}G$ & 7/2  & 5.145($-$07) & 26144.68 \\
 7 & $^{4}P$ & 5/2  & $^{4}G$ & 9/2  & 4.186($-$08) & 26199.60 \\
 8 & $^{4}P$ & 3/2  & $^{6}S$ & 5/2  & 1.951($-$01) &  3218.33 \\
 9 & $^{4}P$ & 3/2  & $^{4}G$ & 5/2  & 3.703($-$07) & 25940.64 \\
10 & $^{4}P$ & 3/2  & $^{4}G$ & 7/2  & 3.068($-$08) & 26011.38 \\
11 & $^{4}P$ & 1/2  & $^{6}S$ & 5/2  & 3.247($-$05) &  3215.61 \\
12 & $^{4}P$ & 1/2  & $^{4}G$ & 5/2  & 1.788($-$08) & 25764.91 \\
13 & $^{4}D$ & 7/2  & $^{6}S$ & 5/2  & 5.270($-$04) &  3000.85 \\
14 & $^{4}D$ & 7/2  & $^{4}G$ & 5/2  & 6.119($-$06) & 16375.30 \\
15 & $^{4}D$ & 7/2  & $^{4}G$ & 7/2  & 5.466($-$05) & 16403.46 \\
16 & $^{4}D$ & 7/2  & $^{4}G$ & 9/2  & 6.268($-$05) & 16425.07 \\
17 & $^{4}D$ & 7/2  & $^{4}G$ & 11/2 & 2.619($-$05) & 16429.05 \\
18 & $^{4}D$ & 7/2  & $^{4}P$ & 5/2  & 6.609($-$03) & 44025.64 \\
19 & $^{4}D$ & 7/2  & $^{4}P$ & 3/2  & 1.682($-$06) & 44408.86 \\
20 & $^{4}D$ & 1/2  & $^{6}S$ & 5/2  & 7.604($-$05) &  2999.80 \\
21 & $^{4}D$ & 1/2  & $^{4}G$ & 5/2  & 3.499($-$05) & 16343.86 \\
22 & $^{4}D$ & 1/2  & $^{4}P$ & 5/2  & 5.838($-$07) & 43799.09 \\
23 & $^{4}D$ & 1/2  & $^{4}P$ & 3/2  & 5.947($-$03) & 44178.36 \\
24 & $^{4}D$ & 1/2  & $^{4}P$ & 1/2  & 9.842($-$03) & 44697.55 \\
25 & $^{4}D$ & 3/2  & $^{6}S$ & 5/2  & 3.601($-$03) &  2997.45 \\
26 & $^{4}D$ & 3/2  & $^{4}G$ & 5/2  & 3.150($-$05) & 16274.34 \\
27 & $^{4}D$ & 3/2  & $^{4}G$ & 7/2  & 2.044($-$05) & 16302.16 \\
28 & $^{4}D$ & 3/2  & $^{4}P$ & 5/2  & 2.998($-$03) & 43303.43 \\
29 & $^{4}D$ & 3/2  & $^{4}P$ & 3/2  & 6.436($-$03) & 43674.12 \\
30 & $^{4}D$ & 3/2  & $^{4}P$ & 1/2  & 2.889($-$04) & 44181.46 \\
31 & $^{4}D$ & 5/2  & $^{6}S$ & 5/2  & 6.818($-$03) &  2996.42 \\
32 & $^{4}D$ & 5/2  & $^{4}G$ & 5/2  & 4.978($-$05) & 16244.05 \\
33 & $^{4}D$ & 5/2  & $^{4}G$ & 7/2  & 1.659($-$05) & 16271.76 \\
34 & $^{4}D$ & 5/2  & $^{4}G$ & 9/2  & 2.007($-$05) & 16293.02 \\
35 & $^{4}D$ & 5/2  & $^{4}P$ & 5/2  & 3.653($-$03) & 43089.58 \\
 \hline
\end{tabular}
\end{table}
\subsection{Collisional excitation}
The collision strengths, $\Omega_{ij}$, for the $E2$ and $M1$ transitions among the fine-structure levels of the ground-state manifold in [Mn\,\textsc{iii}] were calculated using the Breit--Pauli $R$-matrix (BPRM) method, widely employed for accurate and large-scale atomic calculations \citep{pb11,ip,pradhan2011atomic}. The close coupling R-matrix calculations include channel couplings among the 38 energy levels of [Mn\,\textsc{iii}], and the energy range considered extended up to the highest threshold is at 2.2 Ry and above up to 5.0 Ry.
The collision strengths generally exhibit extensive resonance structures, particularly in the vicinity of excitation thresholds, arising from autoionizing levels associated with higher levels included in the close coupling expansion. For astrophysical applications, the plasma electrons are assumed to follow a Maxwellian distribution at a given electron temperature $T_e$, and therefore we compute the Maxwellian averaged effective collision strengths \citep{pradhan2011atomic},
\begin{equation}
\Upsilon_{ij}(T_e) = \int_0^\infty \Omega_{ij}(E)
\exp\!\left(-\frac{E}{kT_e}\right)
\, d\!\left(\frac{E}{kT_e}\right),
\end{equation}
where $\Omega_{ij}(E)$ is the collision strength at incident electron energy $E$, and $k$ is the Boltzmann constant. The effective collision strength, $\Upsilon_{ij}(T_e)$, varies more smoothly and gradually with temperature than the raw collision strength, since the averaging process smooths out the sharp resonance structure. In the present work, $\Upsilon_{ij}$ was calculated at the electron temperatures 2500, 5000, 10000, 20000, 30000 and 40000~K, in order to cover the typical thermal range of astrophysical plasma environments in \hii regions where [Mn\,\textsc{iii}] may be observed.
The collisional excitation rate is related to the effective collision strength through the excitation rate coefficient
\begin{equation}
q_{ij}(T_e) =
\frac{8.629 \times 10^{-6}}{g_i \, T_e^{1/2}}
\, \Upsilon_{ij}(T_e)
\qquad \mathrm{(cm^3\,s^{-1})},
\end{equation}
where $g_i = 2J_i + 1$ is the statistical weight of the initial level $i$.
\vspace{-15pt}

\subsection{Collisional-Radiative Model}
In order to derive the spectral line emissivity ratios as functions
of electron temperature $T_e$ and electron density $N_e$, a
collisional-radiative (CR) model was constructed using the
\textsc{spectra} code \citep{Hoy2023,pradhan2011atomic}. This model
is based on solving the statistical equilibrium equations for coupled level populations coupled via collisional and radiative processes.
The theoretical line emissivity is the amount of energy per unit time per unit
volume for a given transition $j \rightarrow i$, expressed as
\begin{equation}
\begin{aligned}
I([{\rm Mn\,III}];\lambda_{ji}) =
& \frac{h\nu_{ji}A_{ji}}{4\pi}
\frac{N_j}{\sum_k N_k}
\frac{n({\rm Mn\,III})}{n({\rm Mn})} \\
& \times
\frac{n({\rm Mn})}{n({\rm H})}
n({\rm H})
\quad {\rm erg\,cm^{-3}\,s^{-1}} .
\end{aligned}
\label{eq:emissivity}
\end{equation}

\noindent $N_j$ refers to the population of the upper state. 
Summation over $N_k$ refers to the total population of all levels 
included in the CR calculation. The ratio $n$(Mn\,\textsc{iii})/$n$(Mn) is the 
ionization fraction of Mn\,\textsc{iii} relative to total manganese, and 
$n$(Mn)/$n$(H) is the manganese abundance relative to hydrogen deduced from \ha,\hb, etc.
In order to obtain line emissivities from Eq.~(\ref{eq:emissivity}), 
one needs ionization fractions Mn\,\textsc{iii}/Mn in the plasma source at the 
appropriate temperature and density, and are model dependent 
\citep{pradhan2011atomic,Osterbrock2006,dopita2013astrophysics}.

\section{Results and discussion}

Atomic structure and collisional-radiative calculations are carried out for all levels and transitions described above. A brief sample of those results is presented and discussed below.
\subsection{Collision Strengths}
\begin{figure}
    \centering
    \includegraphics[width=\columnwidth]{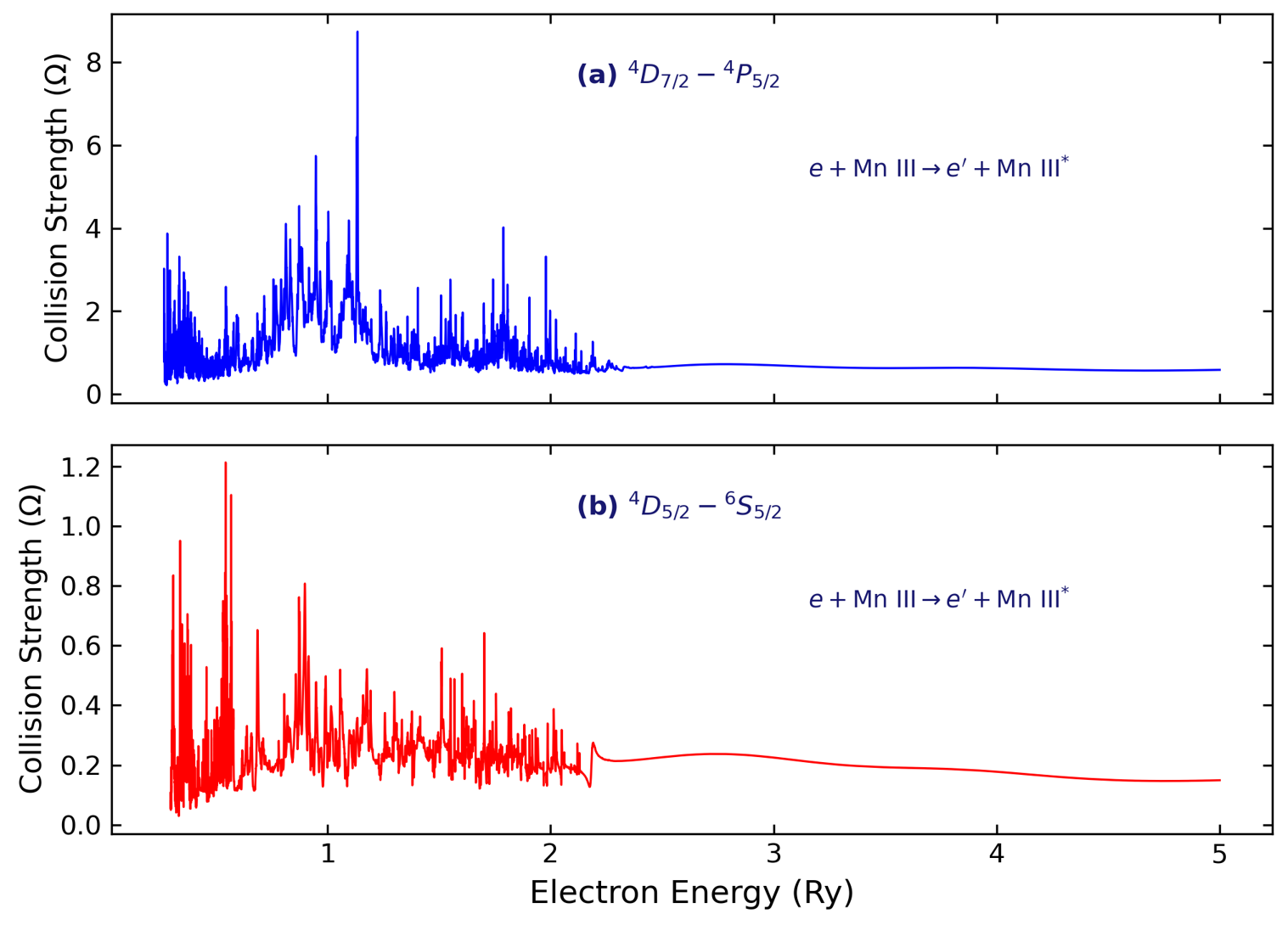}
    \caption{Collision strengths, $\Omega$, for the forbidden transitions
(a) $^{4}D_{7/2} \rightarrow {}^{4}P_{5/2}$ and
(b) $^{4}D_{5/2} \rightarrow {}^{6}S_{5/2}$ in Mn\,III as functions of
electron energy (Ry). Prominent resonance structures are present in
the near-threshold region, while smoother behaviour is observed at
higher energies.}
           \label{fig:omega_panel}
\end{figure}

The collision strengths, $\Omega_{ij}$, for four selected forbidden 
transitions of Mn\,\textsc{iii} are presented in Fig.~\ref{fig:omega_panel} as functions 
of incident electron energy $E$ (Ry), and BPRM collision strengths are obtained for 703 transitions among
38 target levels over the energy range 0.0--5.0~Ry.
All four transitions exhibit resonances in Rydberg series converging on to higher levels included in the 
close coupling expansion up to the highest level at 2.2 Ry, beyond which there is a relatively smooth background.

Figs.~~\ref{fig:omega_panel}a,b show collision strengths for the transitions 
$^4D_{7/2} \rightarrow {^4P_{5/2}}$ ($\lambda = 4.403\,\mu$m) 
and $^4D_{5/2} \rightarrow {^6S_{5/2}}$ ($\lambda = 0.300\,\mu$m), 
respectively, which correspond to the density diagnostic line ratios in Fig.~\ref{fig:density_ratios}.

\subsection{Effective Collision Strengths}

The effective Maxwellian averaged collision strengths (Eq.~1), $\Upsilon_{ij}(T_{\mathrm{e}})$, were obtained over a range of electron temperatures characteristic of \hii regions ($T_{\mathrm{e}} = 2500$--$40000$~K), that are relatively smoothly varying with \te.
For most transitions, the $\Upsilon_{ij}(T_{\mathrm{e}})$ exhibit a 
monotonic variation with temperature, reflecting the enhanced 
contribution of higher-energy electrons to collisional excitation. 
However, for transitions dominated by near-threshold resonances, 
departures from monotonic behaviour may occur at lower temperatures. 
The excitation rate coefficients are computed as in Eq.~(2) in the CR model to obtain level populations.  Calculated values of $\Upsilon_{ij}(T_{\mathrm{e}})$ for selected transitions are listed in Table~3.
%The effective collision strengths, $\Upsilon_{ij}(T_e)$, were obtained by %Maxwellian averaging of the collision strengths $\Omega_{ij}(E)$ over a %range of electron temperatures characteristic of astrophysical plasma %environments ($T_e = 2500$--$30000$~K).
\vspace{-10pt}
\subsection{Line Emissivity Ratios}
When calculating the ratio between the intensities of two spectral
lines as a function of electron density, two distinct behavioural
regimes are identified depending on the energy separation of the
levels involved and on the transition pair considered. For
transitions between levels lying close to the ground state,
characterized by small energy differences and relatively long
wavelengths ($\lambda \sim 3$--$44\,\mu$m), the variation in the
line ratio is noticeable over
$3 \lesssim \log_{10} N_e \lesssim 7$\,cm$^{-3}$. This behaviour
arises because such levels are efficiently populated by
electron-impact excitation even at relatively low electron
densities. As the density increases beyond the critical
density for these transitions, collisional de-excitation
redistributes the level populations, causing the ratio to approach
a density-independent value characteristic of local thermodynamic
equilibrium (LTE) at high densities.

\begin{figure*}
  \centering
  \includegraphics[width=\textwidth]{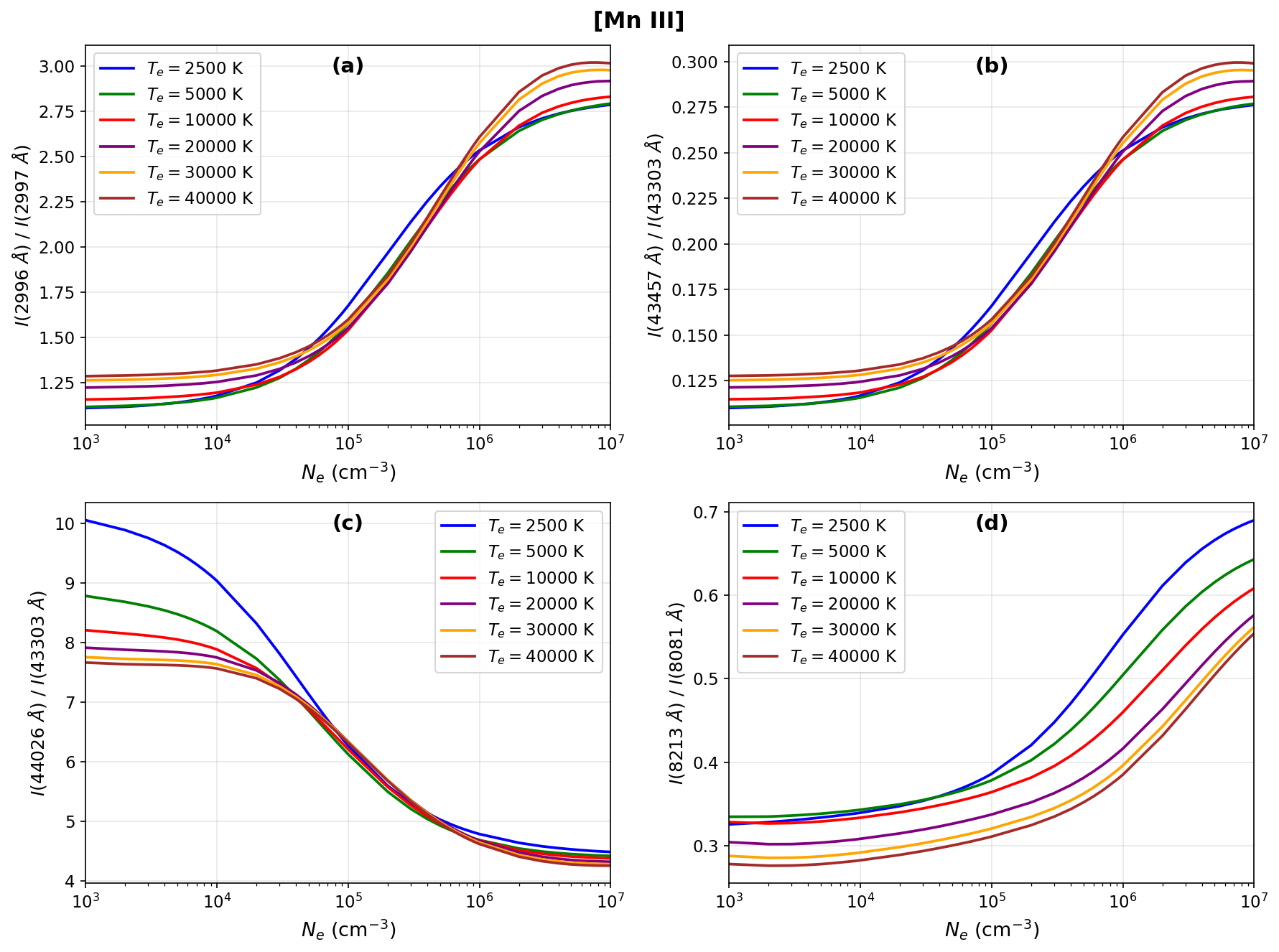}
  \caption{Mn\,{\sc iii} forbidden-line emissivity ratios
  as functions of $\log_{10}(N_{\rm e}/\mathrm{cm}^{-3})$,
  at $T_{\rm e} = 2500$, 5000, 10\,000, 20\,000, 30\,000, and
  40\,000\,K.
  Top row: ratios nearly independent of $T_{\rm e}$:
  (a) $R_1$: $I(2996.42\,\text{\AA})/I(2997.45\,\text{\AA})$,
  $^4D_{5/2}\to{^6S_{5/2}}$\,/\,$^4D_{3/2}\to{^6S_{5/2}}$.
  (b) $R_2$: $I(4.35\,\mu\text{m})/I(4.33\,\mu\text{m})$,
  $^4D_{5/2}\to{^4P_{3/2}}$\,/\,$^4D_{3/2}\to{^4P_{5/2}}$.
  Bottom row: ratios sensitive to both $N_{\rm e}$ and $T_{\rm e}$:
  (c) $R_3$: $I(4.40\,\mu\text{m})/I(4.33\,\mu\text{m})$,
  $^4D_{7/2}\to{^4P_{5/2}}$\,/\,$^4D_{3/2}\to{^4P_{5/2}}$.
  (d) $R_4$: $I(8213.52\,\text{\AA})/I(8081.28\,\text{\AA})$,
  $^2F_{5/2}\to{^4P_{3/2}}$\,/\,$^2D_{3/2}\to{^4P_{3/2}}$.}
  \label{fig:density_ratios}
\end{figure*}

Fig.~\ref{fig:density_ratios} presents four line emissivity ratios
divided into two categories. The top row shows ratios nearly
independent of electron temperature $T_e$, while the bottom row
shows ratios sensitive to both $N_{\rm e}$ and $T_e$.

Panel~(a) shows the ratio $R_1$, defined as the intensity of the
$^4D_{5/2} \rightarrow\, ^6S_{5/2}$ transition at
$\lambda = 2996.42$\,\AA\ to that of the
$^4D_{3/2} \rightarrow\, ^6S_{5/2}$ transition at
$\lambda = 2997.45$\,\AA. Both transitions originate from the
$^4D$ multiplet and share the same lower level $^6S_{5/2}$,
lying within a narrow ultraviolet interval, rendering this ratio
insensitive to interstellar reddening and providing a clean density
diagnostic over $3 \lesssim \log_{10} N_e \lesssim 7$\,cm$^{-3}$,
characteristic of planetary nebulae and \ion{H}{II} regions.

Panel~(b) shows $R_2$, the intensity of
$^4D_{5/2} \rightarrow\, ^4P_{3/2}$ at $4.35\,\mu$m relative to
$^4D_{3/2} \rightarrow\, ^4P_{5/2}$ at $4.33\,\mu$m, which lies
in the mid-infrared and is accessible with JWST.
Both $R_1$ and $R_2$ are virtually independent of $T_e$ over the
range 2500--40\,000\,K, confirming their utility as pure electron
density diagnostics.

Panel~(c) shows $R_3$, the ratio of
$^4D_{7/2} \rightarrow\, ^4P_{5/2}$ at $4.40\,\mu$m to
$^4D_{3/2} \rightarrow\, ^4P_{5/2}$ at $4.33\,\mu$m,
which exhibits a strong dependence on $T_{\rm e}$,
particularly at low densities
($\log_{10} N_{\rm e} \lesssim 4$\,cm$^{-3}$), and is likewise
observable with JWST.

Panel~(d) presents $R_4$, the ratio of
$^2F_{5/2} \rightarrow\, ^4P_{3/2}$ at $8213.52$\,\AA\ to
$^2D_{3/2} \rightarrow\, ^4P_{3/2}$ at $8081.28$\,\AA,
which increases monotonically with $N_{\rm e}$ and shows
significant temperature sensitivity at all densities.
The ratios $R_3$ and $R_4$ may be employed in combination with
$R_1$ and $R_2$ to lift the degeneracy between electron
temperature and density in \hii regions.
The resulting diagnostic line emissivity ratios are presented in Fig.~\ref{fig:density_ratios} and divide into two distinct categories:
\begin{itemize}
  \item \textbf{Pure electron density diagnostics (top row of 
  Fig.~\ref{fig:density_ratios}):} 
  The ultraviolet ratio $R_1$ 
  (panel~a), defined as $I(2996.42\,\text{\AA})/I(2997.45\,\text{\AA})$, 
  involves two transitions within the $^4D$ multiplet sharing the 
  common lower level $^6S_{5/2}$ at nearly identical wavelengths, 
  and is therefore particularly robust against interstellar reddening 
  and instrumental flux calibration uncertainties. The mid-infrared 
  ratio $R_2$ (panel~b), $I(4.35\,\mu\text{m})/I(4.33\,\mu\text{m})$, 
  likewise falls within the observational reach of the James Webb 
  Space Telescope (JWST). Both ratios are virtually insensitive to 
  electron temperature across the full range $T_e = 2500$--$40\,000$\,K, 
  and vary monotonically with electron density over 
  $3 \lesssim \log_{10} N_e \lesssim 7$\,cm$^{-3}$, making them 
  reliable pure density diagnostics for planetary nebulae, 
  \ion{H}{II} regions, and SNRs.
  \item \textbf{Simultaneous temperature--density diagnostics (bottom 
  row of Fig.~\ref{fig:density_ratios}):} The mid-infrared ratio $R_3$ 
  (panel~c), $I(4.40\,\mu\text{m})/I(4.33\,\mu\text{m})$, exhibits a 
  strong dependence on $T_e$, particularly at low densities 
  ($\log_{10} N_e \lesssim 4$\,cm$^{-3}$), and is accessible with JWST. 
  The near-infrared ratio $R_4$ (panel~d), 
  $I(8213.52\,\text{\AA})/I(8081.28\,\text{\AA})$, increases 
  monotonically with $N_e$ and shows significant temperature sensitivity 
  at all densities. These two ratios may be employed in combination 
  with $R_1$ and $R_2$ to break the degeneracy between electron 
  temperature and density in nebular plasma diagnostics.
  
\end{itemize}
%\begin{quote}
%\bfseries
Following extensive analysis of emissivity ratios of all transitions considered in this work, we examine the tentative identification by Esteban et al.\ (2004, Table~2) \cite{Esteban2004} of a Mn~III line at 6821.16~\AA. However, with an uncertainty of, or exceeding, 40 per cent, this identification remains highly uncertain, and we cannot definitively confirm or rule out the presence of a spectral line at this wavelength.
On the one hand, our calculations indicate that the two closest candidate transitions are $^4F_{7/2}\to\,^4P_{3/2}$ at 6801.52~\AA\ and $^4F_{9/2}\to\,^4P_{5/2}$ at 6796.42~\AA, which lie nearest in wavelength to the observed value. We computed the predicted intensity of each of these candidates relative to several strong, well-established reference lines within our spectral range, and found the candidate transitions to be markedly weak: their relative intensity ranged from approximately 26 to 45 times weaker than the $R_4$ diagnostic line ($^2F_{5/2}\to\,^4P_{3/2}$, 8213.52~\AA), and reached its most pronounced weakness -- approximately 370 times weaker -- when compared with a strong, nearby line ($^2F_{7/2}\to\,^4G_{9/2}$, 5747.13~\AA).
Despite this consistent pattern of weakness observed across multiple independent comparisons, we cannot definitively rule out the presence of this line, since our calculations are limited to the relative strength of atomic transitions within the collisional-radiative model, and do not account for other factors that could substantially affect the actual observed intensity. Primary among these are the Mn~III/Mn ionization fraction in the specific nebular environment under study (e.g.\ the Orion nebula) and the Mn/H abundance that are model dependent. It is possible, in principle, that the Mn~III ion concentration is exceptionally low in the region observed, or that other physical factors beyond the scope of the present atomic model are involved, which could account for the observed weakness of the line without necessarily invalidating the proposed spectral identification.
%\end{quote}
%\end{itemize}

\FloatBarrier
\vspace{-20pt}
\section{Conclusions}
We have presented the first systematic theoretical study of forbidden emission lines from doubly ionized manganese, [\ion{Mn}{III}], aimed at establishing physical conditions and abundances in \ion{H}{II} regions, supernova remnants (SNRs) and similar nebular environments. 
A comprehensive set of atomic data was computed for 1421 fine-structure energy levels of \ion{Mn}{III}.
Collision strengths $\Omega_{ij}$ for all 703 forbidden transitions among the lowest 38 fine-structure levels were computed using the Breit--Pauli $R$-matrix (BPRM) method \citep{ip,Berrington1995}. A CR model constructed using the \textsc{spectra} code \citep{Hoy2023,pradhan2011atomic} was employed to compute line emissivity ratios over the electron density range $3 \leq \log_{10} N_e \leq 7$\,cm$^{-3}$. Complete datasets for \mniii will be described in a subsequent paper, and made available from the online database \citealt{norad}.

Employing the atomic data and line emissivities in this work, it should be possible to relate the manganese abundance to that of other elements and overall metallicity in \hii regions. In particular, the Mn/Fe abundance ratio may be derived using JWST observations of high-$z$ galaxies (\eg \citep{Nakane_2025,Sarkar_2025}). Atomic data and line ratios for \feii have been obtained in previous works \citep{pradhan1993,bautista1996}, and may be employed for emission line analysis. In a recent work, we analyzed the temperature-abdundance dependent forbidden [\oiii] line ratios in high-$z$ galaxies with data from JWST-VLT-Keck observations to infer the oxygen abundance and metallicity evolution with redshift, also using the canonical [\oii] and [\sii] forbidden lines as density diagnostics (Bhandari \etal 2026). Coupled with [\feii] emission line diagnostics, the present work on [\mniii] may be employed to determine Mn/Fe ratios in a variety of sources as indicators of the chemical evolution of the Universe.

\vspace{-20pt}
\section*{Acknowledgments}
This work was supported in part by the U.S. National Science Foundation grant AST-2407470. Partial computational work was carried out at the Ohio Supercomputer Center in Columbus, Ohio, USA.

%%%%%%%%%%%%%%%%%%%%%%%%%%%%%%%%%%%%%%%%%%%%%%%%%%
\vspace{-20pt}
\section*{Data Availability}
The atomic data described herein will be made available from the online database NORAD (Nahar-OSU-Radiative-Atomic-Database; \citealt{nahar2020database}), or by correspondence with the authors.

\begin{table}
\centering
\caption{Effective Maxwellian-averaged collision strengths
$\Upsilon_{ij}(T_{\rm e})$ for selected forbidden transitions
of [Mn\,\textsc{iii}] at $T_{\rm e} = 5000$, 10\,000,
and 20\,000\,K. All levels belong to the $3d^5$ configuration.}
\label{tab:upsilon_3T}
\begin{tabular}{l r c c c }
\hline\hline
Transition & $\lambda$ (\AA) & \multicolumn{3}{c}{$\Upsilon(T_{\rm e})$} \\
\cline{3-5}
 & & $5000$ & $10000$ & $20000$ \\
\hline
$^6S_{5/2}$ -- $^4G_{11/2}$ & 3674.16 & $8.582E-01$ & $8.923E-01$ & $9.770E-01$ \\
$^6S_{5/2}$ -- $^4G_{9/2}$ & 3672.75 & $8.202E-01$ & $8.259E-01$ & $8.680E-01$ \\
$^6S_{5/2}$ -- $^4G_{5/2}$ & 3671.67 & $7.029E-01$ & $6.834E-01$ & $7.056E-01$ \\
$^4G_{11/2}$ -- $^4P_{5/2}$ & 26073.21 & $6.476E-01$ & $6.340E-01$ & $6.115E-01$ \\
$^4G_{5/2}$ -- $^4P_{5/2}$ & 26199.60 & $4.938E-01$ & $4.464E-01$ & $3.952E-01$ \\
$^6S_{5/2}$ -- $^4P_{3/2}$ & 3218.33 & $3.696E-01$ & $3.934E-01$ & $3.934E-01$ \\
$^6S_{5/2}$ -- $^4P_{1/2}$ & 3215.61 & $1.819E-01$ & $1.906E-01$ & $1.889E-01$ \\
$^6S_{5/2}$ -- $^4D_{7/2}$ & 3000.85 & $7.084E-01$ & $7.728E-01$ & $7.992E-01$ \\
$^6S_{5/2}$ -- $^4D_{5/2}$ & 2996.42 & $2.136E-01$ & $2.029E-01$ & $1.977E-01$ \\
$^6S_{5/2}$ -- $^4D_{3/2}$ & 2997.45 & $4.459E-01$ & $4.252E-01$ & $4.104E-01$ \\
$^4P_{3/2}$ -- $^4D_{5/2}$ & 43456.61 & $1.667E-01$ & $1.734E-01$ & $1.804E-01$ \\
$^4P_{5/2}$ -- $^4D_{3/2}$ & 43303.43 & $3.300E-01$ & $3.330E-01$ & $3.338E-01$ \\
$^4P_{3/2}$ -- $^4D_{3/2}$ & 43674.12 & $2.758E-01$ & $2.828E-01$ & $2.974E-01$ \\
$^4P_{5/2}$ -- $^4D_{7/2}$ & 44025.64 & $8.761E-01$ & $8.704E-01$ & $8.744E-01$ \\
$^4P_{3/2}$ -- $^4D_{3/2}$ & 43674.12 & $2.758E-01$ & $2.828E-01$ & $2.974E-01$ \\
$^4P_{5/2}$ -- $^4D_{7/2}$ & 44025.64 & $8.761E-01$ & $8.704E-01$ & $8.744E-01$ \\
$^4P_{5/2}$ -- $^4D_{3/2}$ & 43303.43 & $3.300E-01$ & $3.330E-01$ & $3.338E-01$ \\
$^4P_{3/2}$ -- $^2F_{5/2}$ & 8213.52 & $3.670E-01$ & $3.297E-01$ & $3.070E-01$ \\
$^4P_{3/2}$ -- $^2D_{3/2}$ & 8081.28 & $2.946E-01$ & $2.693E-01$ & $2.613E-01$ \\
\hline\hline
\end{tabular}
\end{table}

%========================
% Table A1
%========================

%%%%%%%%%%%%%%%%%%%%%%%%%%%%%%%%%%%%%%%%%%%%%%%%%%

%%%%%%%%%%%%%%%%% APPENDICES %%%%%%%%%%%%%%%%%%%%%


%\clearpage
\vspace{-10pt}
\bibliographystyle{mnras}
\bibliography{example} % if your bibtex file is called example.bib

%\begin{thebibliography}{99}
%\end{thebibliography}

% Don't change these lines
\bsp	% typesetting comment
\label{lastpage}
\end{document}